\magnification=1200
\vsize=8.5truein
\hsize=6truein
\baselineskip=20pt
\centerline{\bf From the Plateau problem to periodic minimal surfaces}
\centerline{\bf in lipids, surfactants and diblock copolymers}
\centerline{by}
\centerline{Wojciech G\`o\`zd\`z and Robert Ho{\l}yst}
\vskip 20pt
\centerline {Institute of Physical Chemistry
PAS and College of Science,}
\centerline {Dept.III, Kasprzaka 44/52,
01224 Warsaw, Poland}

\centerline{\bf Abstract}

The novel method is presented for generating periodic surfaces.
Such periodic surfaces appear in all systems which are
characterized by internal interfaces and which additionally
exhibit ordering. One example is the system of the AB diblock
copolymers, where the internal interfaces are formed by the
chemical bonds between the A and B blocks. In this system at least
two bicontinuous phases are formed :
the ordered bicontinuous double diamond phase and
gyroid phase. In these phases the ordered domains of A monomers and
B monomers are separated by the periodic interface of same symmetry
as the phases themselves.
Here we present the
novel method for the generation of such periodic
surfaces based on the simple Landau-Ginzburg
model of microemulsions. We test the method on four known minimal periodic
surfaces (P,D,G and I-WP),
find two new surfaces of cubic symmetry,
show how to obtain periodic surfaces of high genus  and
n-tuply-continuous phases ((n$>2$)
So far only bicontinuous (n$=2$)
phases have been  known. We point that the Landau model
used here should be generic for all systems characterized by internal
interfaces, including the diblock copolymer systems.

\vfill\eject
\centerline{\bf I. Introduction}

Triply periodic minimal surfaces are paradigm of
surfaces in cubic bicontinuous
phases formed by biological molecules (lipids) and surfactants
in aqueous solutions$^{1-3)}$, silicate membranes
macrostructures$^{4,5)}$,  zero potential surfaces in ionic crystals$^{6)}$,
and domain interfaces in diblock copolymer
structures$^{7ab)}$. They  can be used in the description of structures of
carbon$^{8)}$ ${\bf C_{60}}$,
blue phases of thermotropic liquid crystals$^{9)}$,
blue phases of DNA$^{10)}$, and possibly
cellular structures in the earliest stages of embryogenesis$^{11)}$
and chloroplast structures within plant cells$^{12)}$.
Here we present the
novel method for their generation based on the simple Landau-Ginzburg
model of microemulsions$^{13)}$. We test the method on four known minimal
surfaces (P,D,G and I-WP, Table 1. and Fig.1),
find two new surfaces of cubic symmetry (Figs.(2,3) and Table 1),
show how to obtain periodic surfaces of high genus (Fig.4) and
n-tuply-continuous phases ((n$>2$)
(quadruply-continuous gyroid
structure is shown in Fig.5). So far only bicontinuous (n$=2$)
phases have been  known.

Most biological molecules (e.g. lipids) have both
polar and nonpolar segments and therefore they self-assembly, in water
solutions, in bilayers. These bilayers can further organize in periodic
one, two or three dimensional structures,similarly to soaps
and detergents, as was explicitly demonstrated
by Luzzati et al.$^{1)}$ using x-ray technique.
Although the symmetry of these phases
can in principle be obtained from the x-ray or neutron scattering it is not
clear how to obtain their
geometry and topology from experimental studies alone.
That is why
the periodic
minimal surfaces$^{14-16)}$ have served as paradigm of the
periodic biological (or surfactant) surfaces formed in
solutions. Minimal surface is the surface of zero average curvature at every
point. The fact that such surfaces can form three dimensional periodic
structures was discovered by Schwarz in XIX century.
There are two ways of generating such surfaces: either
from the definition of the average curvature$^{15,16)}$ or from the
Weierstrass parametrization$^{17,18)}$. Both methods are related to the
geometry of the problem and not to the physics of soaps, detergents or
biological molecules in solutions. Here we fill the gap in our understanding
of the connection between physical interfaces and geometrical models
by presenting a simple physical model of microemulsions$^{13)}$
and showing explicitely that physical interfaces in the model are
periodic minimal
surfaces.
As a by-product of our research we have obtained a new method of
generating periodic minimal surfaces. We have tested the method on four
known surfaces i.e. P,D Schwarz and G (Fig.1),I-WP Schoen surfaces and
proved its efficiency by generating new surfaces of cubic symmetry (Table 1
and Figs.2,3).
The knowledge of new types of
surfaces can be useful in indentifying them in real systems.
In the lipid-water system six cubic phases have been found, out of which
only four have been determined unambiguously$^{3)}$.

The structure shown in Fig.5 is of special
interest. It shows three periodic surfaces of nonpositive Gaussian curvature
partitioning the space into four continuous disjoint subvolumes, making
a periodic quadruply continuous structure.
 In all the
examples shown in Figs. (1-4) a single
surface partitions the space into two subvolumes and consequently
the emerging structures are only bicontinuous. The idea of quadruple continuity
or in general n-tuple continuity (n$>2$) has not been previously
considered.

The following simple experiment
can be used for the direct visualization
of a simple patch of minimal surface: Take a metal
non-planar frame and immerse it in the water solution of soap.
The soap bubble which
forms on this frame assumes the shape that minimizes its
surface free energy associated with the surface tension and consequently
it forms a surface of least area.
Thus these surfaces are called minimal surfaces. Such
experiments can be traced back to Leonardo da Vinci$^{19)}$ but in fact the
detailed studies of this type were done and published
by Plateau$^{20)}$ and hence later the
problem of surface of least area spanning a given loop has been named the
Plateau problem.

The history of physics and mathematics of minimal surfaces
ran in parallel. Lagrange in 1761 (before Plateau) derived the  equations
for a surface of least area that is equivalent to the condition
of vanishing mean curvature at every point on the surface.
The representation
of these surfaces in terms of harmonic functions was given
by Weierstrass in 1866 and this representation has served many researchers
up to date for their generations. A new qualitative insight into the
mathematics of the problem was obtained by Schwarz and his student Neovius,
who showed that simple patches of minimal surfaces can be put together to
give smooth periodic three dimensional structures, which are called now
triply periodic minimal surfaces (TPMS) or sometimes infinite
periodic minimal
surfaces (IPMS). They identified five phases three of which were of the
cubic symmetry i.e. P, D and C(P).
Plateau and Schwarz in fact entertained scientific contacts, but none of them
had envisaged the role of these surfaces as physical interfaces.
Rediscovery of the problem is due to
Schoen$^{14)}$,
who identified 4 new surfaces of cubic symmetry(G, I-WP, F-RD and
O,C-TO).
In 1976 Scriven$^{21)}$ observed that such surfaces could be used for the
description of physical interfaces appearing in ternary mixtures of
water, oil and surfactants.
The surfactant molecules, which are the
main ingredient of soaps and detergent have the ability to solubilize
oil in water, two liquids which in the binary mixture at normal conditions
are immiscible. This ability stems from their chemical structures;
a surfactant molecule  has the
polar and nonpolar segments at two ends and thus is simultaneously
hydrophobic and hydrophilic. The term amphiphilic
(from Greek word loving both)
molecule is used
since one end (polar) of the molecule is well solubilized in water while the
other (nonpolar) in oil. Hence the molecule
preferably stays at the interface between
oil and water, forming a monolayer interface.
At high concentration of surfactants the physical interface
made of these molecules orders, forming periodic structures of various
symmetries. In the water solution, without oil,
 the molecules form bilayers instead
of monolayers. In 1967/68 Luzzati et al$^{1,22,23)}$
observed this type of ordering
in the lecithin-water and lipid-water systems. One of the phases
observed by them was the phase of the same symmetry as the G Schoen
minimal surface. It seems that both Schoen`s and Luzzati discovery were
made independently thus we shall use the name Schoen-Luzzati gyroid
phase hereafter (Fig.1).
In fact this phase appears to be very common in biological
systems.
Another example of such surfaces is found in the system
of diblock copolymers, commercially important materials for the production
of plastics. AB diblock copolymer consists of two macromolecules chemicaly
bonded together. At low temperatures the system forms
ordered
A-rich and B-rich
domains, with the points of bondage at the interface between
the domains.
In 1988 Thomas et al$^{7a)}$ observed that
the PS/PI (polystyrene/polyisoprene) diblock copolymer forms a structure
of the same symmetry as the D (diamond) Schwarz surface and argued on the
basis of the relative volume fraction of PS and PI component that the
resulting physical interface must be the surface of constant mean curvature
at every point of the surface.
Such surface belongs to the family of
minimal surfaces$^{14)}$.  Recently
they have also observed the gyroid structure (Fig.1)
in the same system
in the weak segregation regime$^{7b)}$.

Surfaces are ubiquitous.
Even in the ionic crystals one can immagine
a periodic zero potential surfaces
(POPS in short), having same symmetry as the
crystal$^{6)}$.
Although, POPS do not have usually the same geometry
as minimal surfaces (their mean curvature varies along the surface),
nonetheless they share the same
topology (genus etc) and symmetry as the latter.

This short historical survey points to the immense importance of
minimal surfaces in many systems ranging from physics to biology
and chemistry. We think that this is not the last word in the story
of minimal surfaces and their application. Below we present a new method
for their generation.

\centerline{\bf II. The model}

The simple Landau-Ginzburg model considered in this paper has been
proposed by Teubner and Strey$^{24)}$ and Gompper and
Schick$^{13,25)}$
based on the  neutron scattering experiments performed on
microemulsion  (homogeneous ternary
mixture  of oil, water and surfactant) and later experiments and
theory of their wetting properties$^{26,27)}$.
The Landau-Ginzburg free energy functional
has the following form:
$$F[\phi]=\int d^3 r \left(\left\vert\triangle \phi\right\vert^2+
\left(g_2\phi^2-g_0\right)
\left\vert\nabla \phi\right\vert^2+(\phi^2-1)^2(\phi^2+f_0)
\right)\eqno(1)$$
where $\phi$, the order parameter, has the interpretation of the
normalized
difference between oil and water concentrations; $g_2$, $g_0$ are two
positive constants and $f_0$ can be of either sign. The sign of the
latter depends on the
stability of bulk microemulsion phase (average order parameter is zero):
for $f_0>0$ microemulsion is a metastable bulk phase
whereas pure water phase ($\phi=1$)
or pure oil phase ($\phi=-1$) are stable; for $f_0\le 0$ microemulsion is
stable. For $g_0>2$
the system can undergo a transition to periodically ordered phases
where water rich domains and oil rich domains order similarly as in the
case of the aforementioned AB
diblock copolymer system$^{7ab)}$. The interface between the
domains corresponds to $\phi ({\bf r})=0$. The only
stable ordered structure is the
lamellar phase, however there exists a large number of ordered metastable
phases, corresponding to the local minima of the functional (1).
It is the main result of this paper  that in many of these
metastable phases the
physical interface between the domains, given by $\phi ({\bf r})=0$, is a
triply periodic minimal embeded surface. We have obtained  6 cubic structures
listed in Table 1 including known examples of P, D, G (Fig.1),
I-WP and
and two new phases shown in Figs. (2,3). The high genus surface shown in Fig.4
has nonpositive Gaussian curvature and large internal surface area,
but numerical accuracy does not allow
us to claim that it is a minimal surface.

In order to find the local minima of the functional we have discretized Eq(1)
on the cubic lattice.
Thus the functional $F[\phi({\bf r})]$ becomes a function
$F(\{\phi_{i,j,k}\})$ of $N^3$
variables, where $Nh$ is the linear dimension of
the cubic lattice and $h$ is the distance between the lattice points.
Each variable $\phi_{i,j,k}$ represents
the value of the field $\phi({\bf r})$ at the lattice site $S=(i,j,k)$,
and the indices $i,j,k$ change from 1 to N. In our calculations we use N=129,
which results in over 2 milion points per unit cell.  The first and
second derivatives in the gradient and laplasian term of the functional (1)
were calculated on the lattice according to the following formulas$^{28)}$
$$g({\bf r}){\partial\phi ({\bf r})
\over{\partial x }} \to
g_{i,j,k}{\phi_{i+1,j,k} - \phi_{i-1,j,k} \over{ 2 h }}\eqno(2) $$
and
$$
{\partial^2\phi ({\bf r})
\over{\partial x^{2} }} \to
{1\over{12h^2}}\left( - \phi_{i+2,j,k} +16 \phi_{i+1,j,k}  -30\phi_{i,j,k}
+ 16\phi_{i-1,j,k}- \phi_{i-2,j,k}
\right)\eqno(3)
$$
 and similar in $y$ and $z$ directions.

We impose on the field $\phi ({\bf r} ) $  the symmetry of the structure,
we are looking for, by building up the field inside a unit
cubic cell from a smaller polyhedron, replicating it by
reflections.
For example structures having $m{\bar 3}m$ space group symmetry are build of
quadrirectangular tetrahedron cut out of the unit cubic
cell by the planes of symmetry. The polyhedrons that we have used
to construct the cubic unit cells are the same as those
described by Coxeter as kaleidoscopic cells$^{14,15)}$.
Such a procedure enables substantial
reduction of independent variables $\phi_{i,j,k}$ in the function
$F(\{\phi_{i,j,k}\})$. We impose on the field $\phi ({\bf r} )$ the periodic
boundary conditions
in $x$ , $y$ and $z$ directions.

The topology of the structure is set up by building the field
$\phi ({\bf r})$ first on a small lattice $N=3$ or 5 analogicaly to
a two component (A,B) molecular crystal. The value of the field
$\phi_{i,j,k}$ at a lattice site $S=(i,j,k)$ is set to 1 if in
the molecular crystal an atom A is in this place, if there is
an atom B $\phi_{i,j,k}$ is set to -1, if there is an empty place
$\phi_{i,j,k}$ is set to 0. Next the small lattice can be enlarged
to desired size by changing the number of points from
$N$ to $2N-1$ and finding the values of $\phi_{i,j,k}$
in new lattice sites by interpolation.

We have used the
conjugate gradient method$^{29)}$ to find a minimum of the function
$F(\{\phi_{i,j,k}\})$.
It is highly unlikely, because of numerical accuracy,
that a value of the field $\phi_{i,j,k}$
at a latice site $S=(i,j,k)$ is zero.
Therefore the points of
the surface have to be localized by linear interpolation between the neighbour
sites of the lattice. This approximation is legible
because the field $\phi({\bf r})$ is very smooth.
The structures of simple topology are formed for small sizes of a unit
cell. The length of a unit cell, $d$, has to be increased to obtain the
structures of complex topology.

In order to calculate genera of the surfaces we use the following method.
For every point $P=(i,j,k)$ where $i \in [1,N-1] $ ,
$j \in [1,N-1] $ , $k \in [1,N-1] $ we examine a cube
formed by this point and vectors ${\bf e}_1 = [ h,0,0] $ ,
${\bf e}_2 = [ 0,h,0] $ , ${\bf e}_3 = [ 0,0,h] $.
If the values of the field at all vertices of this cube are not
the same sign the surface $\phi({\bf r})=0$ must lie inside this
cube. Therefore points of the surface lie on the edges of the cube.
These points form a polygon: a triangle , a tetragon, a pentagon, or
a hexagon. There are no other possibilities. The edges of this
polygon must lie on the faces of the cube, the vertices on the edges
of the cube. Therefore each edge of the polygon belongs to two cubes
and each vertex to four cubes. Thus we can calculate the
number of edges E , vertices V , and faces F of the surface inside a
cubic unit cell by summing F, E, V in all small cubes and taking the
number of vertices with a weight $1/4$ and the number of edges
with a weight $1/2$.  Next the Euler characteristic $\chi$ can be
calculated from the formula
$\chi = F+V-E$ and genera $g$ of the structures from $g = 1 - \chi / 2 $.

 From the form of Eq(1) one can realize that indeed
for some local minima of (1) the average curvature
given by$^{30,31}$:
$$H=-{1\over 2}
\nabla\left({{\nabla\phi}\over{\vert\nabla\phi\vert}}\right)=
-{1\over
2}{{\triangle\phi}\over{\vert\nabla\phi\vert}}+{{\nabla\phi\nabla\vert\nabla\phi
\vert}\over{2\vert\nabla\phi\vert^2}}\eqno(4)$$
vanishes at every point of the $\phi({\bf r})=0$ surface. It follows
from the second term of Eq(1) that $\vert\nabla\phi\vert$ should have the
maximal value  for
$\phi({\bf r})=0$ and consequently the second term
(which after a small algebra can be written as
$(\partial\vert\nabla\phi\vert /\partial n)/2\vert\nabla\phi\vert$, with
$\partial n$ denoting the derivative along the normal to the surface)
 in Eq(4) vanishes. Also
for the $\phi$, $-\phi$ symmetry we know that $H$ averaged over the whole
surface should be zero. It means that either $\triangle\phi$ is exactly zero
at the surface or it changes sign. From the first term of Eq(1) it
follows that the former is favored and consequently $H=0$ at every
point at the surface and hence the surface is minimal.
We have checked numerically that indeed the surface $\phi({\bf r})=0$
coincides with  $H=0$ in all the cases listed in Table 1.

\centerline{\bf III. Summary}

We have presented the novel method for the generation of
periodic surfaces. The method should be useful for
mathematicians working in topology and geometry of surfaces,
crystallographers studying self assembling soft matter
systems, physicists and biologists. The method can also find
application in the design of well characterized mesoporous materials.
For example in the production of such materials the internal interface
in the surfactant system is used as a template for the three dimensional
polimerization of silicate. One obtains in such a way the ordered silicate
pore structure characterized by the same symmetry, topology and geometry
as the surfactant template.

Surfaces are also important for biology.
We finish this report by giving a few examples, not commonly
known, of the influence of
the physical structure on the functions of biological systems.
The concentrated solutions of DNA$^{32-36)}$,
po\-lypeptides$^{36)}$ and polysaccharides$^{36)}$ form ordered
phases. It has been
also observed that DNA in bacteriophages and sperm nuclei of
sepia, trout and salmon
exhibit ordering$^{34)}$. Moreover DNA has been shown to
form blue phases$^{10)}$.
The  activity of DNA  (renaturation, transcription or
replication) can be enhanced in the condensed  phase$^{34)}$.
For example it has been observed that in the phenyl, water
emulsion the renaturation rate can be increased 1000 times in comparison to the
renaturation in standard conditions$^{37)}$.
The renaturations most likely
takes place at the interface between the two phases where the  local
density of DNA strands can be greatly increased. Despite of the
accumulation of great body of experimental results
the problem of
how the geometry
and symmetry of surfaces can affect the biological processes
remains still an open problem.

Let us at the end point out that the Landau-Ginzburg free energy
which we have used in this paper should be generic for all the
systems characterized by the internal interfaces. The same
free energy has been obtained in the ternary mixture
of A homopolymers, B homopolymers and AB diblock copolymers$^{38)}$.

\centerline{\bf Acknowledgements}

This work was supported by two grants from Komitet
Bada\'n Naukowych and a grant from Fundacja Wsp\'o\l pracy
Polsko-Niemieckiej.

\vfill\eject

\centerline{\bf References}
\item{$^{1)}$} V. Luzzati, T. Gulik-Krzywicki, A. Tardieu, A., {\it Nature}
{\bf 218}, 1031-1034 (1968).
\item{$^{2)}$} W. Longley and T.J. McIntosh,
{\it Nature} {\bf 303}, 612-614
(1983).
\item{$^{3)}$} T. Landh, {\it J.Phys.Chem.} {\bf 98}, 8453-8467 (1994).
\item{$^{4)}$} A.L. Mackay, {\it 5th Eur.Crystallogr. Meet., Collected
Abstracts},
73-PI-5a, Co\-pen\-ha\-gen 1979.
\item{$^{5)}$} A. Monnier et al, {\it Science}, 1299-1303 (1993).
\item{$^{6)}$} S. Andersson, Angew.Chem. Int.Ed.Engl. {\bf 22}, 69-170 (1983);
H.G. Schnering and R. Nesper, {\it Angew. Chem.} {\bf 28},
1059-1200 (1987).
\item{$^{7a)}$} E.L. Thomas, D.M. Anderson, C.S. Henkee, D. Hoffman,
{\it Nature} {\bf 334}, 598-601 (1988).
\item{$^{7b)}$} D.A. Hajduk et al, {\it Macromolecules} {\bf 27},
4063-4075 (1994).
\item{$^{8)}$} H. Terrones and A.L. Mackay,
{\it Chem.Phys.Lett.} {\bf 207}, 45-50
(1993).
\item{$^{9)}$} B. Pansu, and E. Dubois-Violette,
{\it J. de Physique Colloque}
{\bf 51}, C7 281-296 (1990).
\item{$^{10)}$}
A. Leforestier and F. Livolant, {\it Liquid Crystals}, {\bf 17},
651-658 (1994).
\item{$^{11)}$} A.L. Mackay, {\it J. de Physique Colloque}, {\bf 51} C7,
399-405
(1990).
\item{$^{12)}$}
A.L. Mackay and J. Klinowski, {\it Comp. \& Maths with Appl.}
{\bf 128}, 803-824 (1986);
S.Lidin, S.T. Hyde, and B.W. Ninham, {\it J. de Physique}
{\bf 51}, 801-813 (1990).
\item{$^{13)}$}
G.Gompper and M.Schick, {\it Self Assembling Amphiphilic Systems},
vol. {\bf 16} {\it Phase Transitions and Critical Phenomena} eds. C.Domb and
J.L.Lebowitz, Academic Press (1994).
\item{$^{14)}$} A.H. Schoen, {\it NASA Technical Note D-5541},
Washington D.C., USA (1970).
\item{$^{15)}$} D.M. Anderson, H.T. Davis, J.C.C. Nitsche,
L.E. Scriven,
{\it Adv.Chem.Phys.} {\bf 77}, 337-396 (1990).
\item{$^{16)}$} A.L. Mackay, {\it Nature} {\bf 314}, 604-606 (1985).
\item{$^{17)}$} H. Terrones,
{\it J. de Physique Colloque} {\bf 51}, C7 345-362
(1990).
\item{$^{18)}$} S.T. Hyde, {\it Z. Kristallogr.} {\bf 187},
165-185 (1989);
E. Koch and W. Fischer, {\it Acta Cryst.} A{\bf 46}, 33-40 (1990);
D. Cvijovi\'c and J. Klinowski, {\it Chem.Phys.Lett.} {\bf 226},
93-99 (1994); {\it J.Phys. I France}, {\bf 3}, 909-924 (1993).
\item{$^{19)}$} J.C. Maxwell and Lord Rayleigh,
in {\it Encyclopedia Brytanica},
13th ed., vol {\bf 5}, 256-275; A.W. Porter,
{\it ibid}, {\bf 21}, 595 (1964).
\item{$^{20)}$} J.A.F. Plateau,
{\it Statique Exp\'erimentale et Th\'eorique des
Liquides Soumis aux Seules Forces Mol\'eculaires} (Gauthier-Villars,
Trubner et Cie, F.Clemm) vol {\bf 2} (1873).
\item{$^{21)}$} L.E. Scriven, {\it Nature} {\bf 263}, 123-125 (1976).
\item{$^{22)}$} V. Luzzati, P.A. Spegt,
{\it Nature} {\bf 215}, 701-704 (1967).
\item{$^{23)}$} V. Luzzati, A. Tardieu, T. Gulik-Krzywicki,
{\it Nature} {\bf 217}, 1028-1030 (1968).
\item{$^{24)}$} M. Teubner and R. Strey,
{\it J.Chem.Phys.} {\bf 87}, 3195-3200 (1987).
\item{$^{25)}$} G. Gompper and M. Schick,
{\it Phys.Rev.Lett} {\bf 65}, 1116-1119
(1990).
\item{$^{26)}$} K.-V. Schubert and R. Strey
{\it J.Chem.Phys.} {\bf 95},
8532-8545 (1991).
\item{$^{27)}$} G. Gompper, R. Ho\l yst and M. Schick,
{\it Phys.Rev. A}
{\bf 43}, 3157-3160 (1991); J. Putz, R. Ho\l yst and M. Schick,
{\it ibid} {\bf 46}, 3369-3372 (1992);
{\it Phys.Rev. E} {\bf 48}, 635 (1993).
\item{$^{28)}$}
{\it Handbook of Mathematical Functions With Formulas, Graphs and
Mathematical Tables} eds. Abramowitz and Stegun,
NBS Applied Mathematics Series
{\bf 55} 883-885 (1964).
\item{$^{29)}$} W.H., Flannery, B.P., Teukolsky, S.A., and Vetterling,
W.T., {\it Nu\-me\-ri\-cal Re\-ci\-pes} , pp 301- 307 , (1989)
\item{$^{30)}$} I.S.Barnes, S.T. Hyde, and B.W. Ninham,
{\it J. de Physique Colloque}
{\bf 51} C7, 19-24 (1990).
\item{$^{31)}$}
M.Spivak {\it A Comprehensive Introduction to Differential
Geometry} vol {\bf III } Publish or Perish Berkley 202-204 (1979).
\item{$^{32)}$} Feughelman et al, Nature {\bf 175}, 834-836 (1955)
\item{$^{33)}$}
F.Livolant, A.M.Levelut, J.Doucet and J.P.Benoit, Nature {\bf 339},
724-726 (1986)
\item{$^{34)}$} F.Livolant, J.Mol.Biol. {\bf 218}, 165-181 (1991);
F.Livolant Physica A
{\bf 176}, 117-137 (1991)
\item{$^{35)}$} D.Durand, J.Doucet and F.Livolant,
J.Phys II {\bf 2}, 1769-1783 (1992).
\item{$^{36)}$} F.Livolant and Y.Bouligand, J.Physique {\bf 48}, 1813-1827
(1986).
\item{$^{37)}$} J.-L. Sikorav and G.M. Church {\it J.Mol.Biol.} {\bf 222},
1085-1108 (1991).
\item{$^{38)}$} R.Ho\l yst and M.Schick, {\it J.Chem.Phys.},
{\bf 96}, 7728-7738 (1992); M.W.Matsen and M.Schick, {\it Macromolecules}
{\bf 26}, 3878-3884 (1993).

\vfill\eject
\centerline{\bf Table Caption}
\item{Table 1.} The minimal surfaces obtained using our method.
In the second column the symmetry is given. We have assumed that two
sides of the surface are equivalent.
The volume fraction, $V/V_{\rm tot}$ (third column)
is the volume at one side of the surface, $V$,
divided by the total volume of the unit cell, $V_{\rm tot}=d^3$, where $d$
is the dimensionless linear size of the unit cell.
The error of the calculation of the volume fraction
has been smaller than 0.001 for most
structures.
The surface area, $S$ (fourth column) is divided by $V^{2/3}=d^2$ i.e. is
calculated per side of the unit cubic cell. The genus, $g$,
has been calculated from
the formula $1-\chi/2$, where $\chi$ is the Euler characteristic per unit cell.
All these data are model independent. The model dependent parameter is the
size of the unit cell $d$ and therefore is not listed in the Table; here we
find for P surface $d=7.88$,
for D, $d=12.56$, for G $d=10.08$ for I-WP $d=11.78$ for
BFY ("butterfly'') $d=16.70$ and for CPD $d=15.12$ (obtained for
$g_0=3$, $f_0=0$ and $g_2=4\sqrt{1+f_0}+g_0+ 0.01$).
All these structures were
obtained for many different
values of $g_2$, $g_0$ and $f_0$. In all cases the parameters
shown in the Table
were the same, although $d$ was different. For P and G surfaces the  worst
error of the  average curvature was 0.01
(typical Gaussian curvature was -0.3), but at most points the average
curvature was smaller than 0.005.
In the D and I-WP surface the error was smaller than
0.015, whereas for the more
complicated structures (BFY and CPD) it was smaller than 0.05.
Because of the errors
we cannot claim with certainty that BFY and CPD phases
are minimal.
In the Table we also
give the  {\bf exact}
values known from literature (if available) for comparison
(shown in parenthesis with the reference number).
The calculations
were performed on the workstation IBM RS/6000 3BT and the figures were
prepared using Data Explorer.

\vfill\eject
\centerline{\bf Figure Captions}
\item{Fig.1} The Schoen-Luzzatti gyroid, G, structure.
One unit cell is shown.
The white
surface (gyroid minimal surface) located at $\phi ({\bf r})=0$ (see Eq(1))
corresponds to the surface dividing oil ($\phi <0$)
and water ($\phi >0$) channels into disjoint
subvolumes. This phase has been recently observed in the diblock
copolymer system$^{7b)}$.
\item{Fig.2} The new (presumably)
minimal surface BFY (Table 1). Legend as in Fig.1.
({\bf a}) 1/8 of the unit cell. ({\bf b}) One unit cell.
\item{Fig.3} The new (presumably)
minimal surface CPD (Table 1). Legend as in Fig.2
({\bf a}) 1/8 of the unit cell. ({\bf b}) One unit cell.
\item{Fig.4} The new structure of cubic symmetry
(same as D surface Table 1.) and very high genus $g=73$.
The legend as in Fig.1.
The surface area $S/V^{2/3}=8.25\pm 0.05$, $V/d^3=0.5$,
 $d=28.88$ (generated at the same values of $g_0$
$f_0$ and $g_2$ as in Table 1).
The surface has nonpositive Gaussian curvature
at every point of the surface. It is possible that it is also a
minimal surface.
({\bf a}) 1/8 of the unit cell. ({\bf b})
One unit cell.
\item{Fig.5} The quadruply continuous cubic structure of the gyroid
symmetry (Table 1). Here there are four
disjoint subvolumes (two "water'' channels and two "oil''
channels) separated by
three surfaces. The middle surface is the G minimal surface (see Fig.1),
whereas
the other two are not. All the surfaces have nonpositive Gaussian curvature
at every point.
Each of the three surfaces has the same genus per unit cell
as the G phase (Table 1).
The surface area $S/V^{2/3}=7.55\pm 0.05$, $V/d^3=0.5$,
 $d=26.28$ ($g_0$
$f_0$ and $g_2$ same as in Table 1).
({\bf a}) 1/8 of the unit cell. ({\bf b})
One unit cell.
\vfill\eject
\centerline{\bf TABLE 1}

$$\vbox{
\hrule  height0.03in
\settabs 5 \columns
\+ \cr
\+ {\bf Name} & {\bf Symmetry} & {\bf Volume   } &{\bf Surface} &{\bf Genus }
\cr
\+       &               & {\bf Fraction}  &      {\bf Area }
&    \cr
\+ \cr
\hrule height0.03in
\+ \cr
\+  P    & $Im\bar{3}m$  & 0.5        & $ 2.3453             $    & $3$    \cr
\+       &               &            & $ [2.3451068]^{(14,15)}   $    &
$[3]^{(14,15)}$    \cr
\+ \cr
\hrule  height0.015in
\+ \cr
\+  D    & $Pn\bar{3}m$  & 0.5        & $ 3.8387             $    & $9$    \cr
\+       &               &            & $ [3.8377862]^{(15)}  $    &        \cr
\+ \cr
\hrule height0.015in
\+ \cr
\+  I-WP & $Im\bar{3}m$  & 0.533      & $ 3.4640             $    & $7$    \cr
\+       &               &            & $ [3.4646016]^{(18)}   $    &
\cr
\+ \cr
\hrule  height0.015in
\+ \cr
\+  G    & $Ia\bar{3}d$  & 0.5        & $ 3.0919             $    & $5$    \cr
\+ \cr
\hrule height0.015in
\+ \cr
\+  CPD  & $Im\bar{3}m$  & 0.515      & $ 4.3588             $    & $14$   \cr
\+ \cr
\hrule  height0.015in
\+ \cr
\+  BFY  & $Im\bar{3}m$  & 0.5        & $ 4.9641             $    & $19$   \cr
\+ \cr
\hrule height0.03in
\+ \cr
} $$

\vfill\eject\end